\documentclass[preprint]{aastex}
\usepackage{natbib}
\usepackage{booktabs}
\usepackage{threeparttable}
\usepackage{multirow}
 
\newcommand{\cii}{C \scriptsize{II} \normalsize}
\newcommand{\civ}{C \scriptsize{IV} \normalsize}
\newcommand{\heii}{He \scriptsize{II} \normalsize}
\newcommand{\siiv}{Si \scriptsize{IV} \normalsize}

\newcommand{\mgii}{Mg \scriptsize{II} \normalsize}
\newcommand{\feviii}{Fe \scriptsize{VIII} \normalsize}
\newcommand{\feix}{Fe \scriptsize{IX} \normalsize}
\newcommand{\fexii}{Fe \scriptsize{XII} \normalsize}
\newcommand{\fexv}{Fe \scriptsize{XV} \normalsize}
\newcommand{\fexxiv}{Fe \scriptsize{XXIV} \normalsize}
\newcommand{\fexiv}{Fe \scriptsize{XIV} \normalsize}
\newcommand{\fexvi}{Fe \scriptsize{XVI} \normalsize}
\newcommand{\fexviii}{Fe \scriptsize{XVIII} \normalsize}
\newcommand{\fexxi}{Fe \scriptsize{XXI} \normalsize}

\begin{document}

\title{Spectroscopic and Stereoscopic Observations of the Solar Jets}

\author{Lei~Lu\altaffilmark{1}, Li~Feng\altaffilmark{1}, Ying~Li\altaffilmark{1}, Dong~Li\altaffilmark{1,2},  Zongjun~Ning\altaffilmark{1}, and Weiqun~Gan\altaffilmark{1}}

\affil{$^1$Key Laboratory of Dark Matter and Space Astronomy, Purple Mountain Observatory, CAS, Nanjing 210033, People's Republic of China \\
       $^2$State Key Laboratory of Space Weather, Chinese Academy of Sciences, Beijing 100190, People's Republic of China}
\altaffiltext{1}{Correspondence should be sent to: lidong@pmo.ac.cn}

\begin{abstract}

We present a comprehensive study of a series of recurrent jets that occurred at the periphery of the NOAA active region 12114 on 2014 July 7.
These jets were found to share the same source region and exhibited rotational motions as they propagated outward.
The multi-wavelength imaging observations made by the AIA and {\it IRIS} telescopes reveal that some of the jets contain cool plasma only, while some others contain not only cool but also hot plasma.
The Doppler velocities calculated from the {\it IRIS} spectra show a continuous evolution from blue to red shifts as the jet motions change from upward to downward. 
Additionally, some jets exhibit opposite Doppler shifts on their both sides, indicative of rotating motions along their axes.
The inclination angle and three-dimensional velocity of the largest jet were inferred from the imaging and spectroscopic observations, which show a high consistence with those derived from the stereoscopic analysis using dual-perspective observations by {\it SDO}/AIA and {\it STEREO}-B/EUVI. 
By relating the jets to the local UV/EUV and full-disk {\it GOES} X-ray emission enhancements, we found that the previous five small-scale jets were triggered by five bright points while the last/largest one was triggered by a C1.6 solar flare. 
Together with a number of type III radio bursts generated during the jet eruptions as well as a weak CME that was observed in association with the last jet, our observations provide evidences in support of multi-scale magnetic reconnection processes being responsible for the production of jet events. 

\end{abstract}


\section{Introduction}

Solar jets are one of the significant transient features occurring on the Sun, which might be an important energy and mass source that maintains the hot corona and the outward-flowing solar wind, especially in the solar quiet region and coronal hole regions. 
Usually, they are magnetic eruptions along vertical or oblique magnetic field lines \citep{Liu11,Shen12,Young14},
while in some cases, they can also occur in a fan-spine magnetic topology \citep{Masson09,Shen19}.
Solar jets share a lot of common properties with some major solar eruptions, such as flares \citep{Zhang14a}, prominences/filaments \citep{Wang18}, and coronal mass ejections \citep[CMEs,][]{Innes13}, which are usually explosive, magnetic-driven, and highly dynamical \citep[see a review by][]{Raouafi16}.
Therefore, the studies of solar jets could be helpful to understand the potential drivers of these larger and more complex solar activities \citep{Tian12,Sterling15,Raouafi16}. Furthermore, solar jets could also provide a critical insight into the basic processes of solar physics, such as particle acceleration and coronal heating \citep[e.g.,][]{Brueckner83,Cirtain07,Dep11,Liu14}.

Observations show that solar jets usually have columnar structures, root in photosphere, and shoot up into the corona or external solar atmospheres \citep{Moore10,Pucci13}, and can cover a broad range (temperature coverage, scale size, and velocity range, etc.) of plasma ejection events, such as X-ray jets \citep{Shibata92,Yokoyama95,Yang11}, extreme-ultraviolet/ultraviolet (EUV/UV) jets \citep{Liu11,Zhang14b}, chromospheric jets \citep{Liu09,Morton12}, network jets \citep{Tian14,Kayshap18}, EUV macrospicules \citep{Bohlin75,Kiss17}, type-II spicules \citep{Dep07,Tian11}, and H$\alpha$ surges/macrospicules \citep{Canfield96,Chae99}. The temperature of these ejected plasmas usually ranges from $\sim$10$^4$~K to $\sim$10$^6$~K \citep{Shimojo96,Zhang16}. The X-ray or EUV jets are usually observed at a temperature of more than 10$^5$~K with a typical height/length of around 10$-$400~Mm and width of about 5$-$100~Mm. The number density in these jets is estimated to be $\sim$(0.8$-$5)$\times$10$^{10}$~cm$^{-3}$ and the apparent velocities usually vary between 10$-$1000~km~s$^{-1}$ with an average value of $\sim$200~km~s$^{-1}$ \citep{Shimojo96,Feng12,Young14,Zhang14a,Raouafi16}. The observed chromospheric jets or H$\alpha$ surges/macrospicules, on the other hand, often exhibit a lower temperature ($\sim$10$^4$~K) and a smaller size, i.e., a short length and a narrow width \citep{Tsiropoula12}.
Another type of solar jets, termed two-sided-loop jets, were identified in X-ray images. They are ejected from the reconnection site between the  emerging flux and the pre-existing horizontal magnetic field lines along two opposite directions \citep{Shibata94, Yokoyama95, Tian17, Zheng18, Shen19b}.
These jet-like events might be closely related to each other, however, the exact relationship still remains unclear and more evidence needs to be found \citep{Morton12}.

With the increasing temporal and spatial resolution of the current imaging and spectroscopic instruments, such as the {\it Solar Terrestrial Relations Observatory} ({\it STEREO}), the {\it Solar Dynamics Observatory} ({\it SDO}), {\it Hinode}, and the {\it Interface Region Imaging Spectrograph} ({\it IRIS}), the fine structures and dynamical behaviours of solar jets have been studied in more and more details and many new results have been obtained. For example, the cool and hot components in solar jets are observed with several minutes delay \citep[see][]{Nishizuka08,Shen12,Mulay17,Shen17}. Some EUV blobs are found in the recurrent/homologous jets, which are explained as the plasmoids produced by magnetic reconnection and ejected outward along the magnetic field lines\citep{Zhang14b,Zhang16,Ni17,Shen17,Zhang19}. The quasi-periodic intensity oscillations are found to be associated with solar jets \citep{Morton12,Zhang14a}. The Kelvin-Helmholtz instability caused by strong-shear flows is observed at the boundaries of solar jets \citep{Bogdanova18,Li18,Zhao18}. These observational results indicate the multi-temperature structures and high dynamics in solar jets. The solar jets often rotate or swirl along their axes, which can be clearly seen in the Doppler shifts of spectral lines \citep{Kamio10,Cheung15,Kayshap18} or the time-distance plot along a slit perpendicular to the jet axis \citep{Shen11a,Zhang14a,Liu14}. The helical structures and untwisting motions of solar jets are also reported, indicating the magnetic-untwisting waves in solar jets \citep{Hong13,Moore15}. 

A large number of observations indicate that the jets are probably driven by magnetic reconnection, which are usually associated with new emerging fluxes or flux cancelations between opposite polarities \citep[e.g.,][]{Yokoyama96,Chifor08,Pontin13,Chen16,Tian18}. Thus, they are often detected in the coronal holes \citep{Moreno08,Adams14,Panesar18}, near the edge of active regions \citep{Kim07,Guo13,Zheng18}, and on the solar limb \citep{Nistico09,Shen11a,Joshi18}, where open magnetic field lines usually dominate. On the other hand, solar flares and filament eruptions are often thought to be the important manifestations of magnetic reconnection at various scales \citep{Solanki93,Priest02,Mackay10}, and many energetic particles could be accelerated and ejected into the outer corona and interplanetary space along the open magnetic field lines during the reconnection process, causing the so-called type III radio burst.

Numerous kinds of solar jets and their relationship to some other solar eruptions (such as large-scale filament oscillations, coronal waves, and coronal mass ejections, etc.) have been reported \citep[e.g.,][]{Luna14,Chandra2015,Shen18a, Shen18b}.
However, the physical nature of solar jets is still an open question, and a one-to-one correspondence in observations at different wavelengths is still lacking. In this paper, making use of the observations from the Atmospheric Imaging Assembly \citep[AIA,][]{Lemen12} on board {\it SDO}, {\it IRIS} \citep{Dep14}, Extreme-Ultraviolet Imager (EUVI) on board {\it STEREO}-Behind \citep[{\it STEREO}-B,][]{Kaiser08}, and {\it Wind}/WAVES \citep{Bougeret95}, we present a detailed investigation of a series of recurring and rotating jets occurred on the south-east solar limb.
Observations at different wavelengths could imply different aspects of the same physical process. By combining them together, we aim to reveal the overall physical picture of the solar jets and their relationships with other eruptions on the Sun.

\section{Observations}

The solar jets under study occurred on 2014 July 7, which were captured by multiple instruments.  
This section describes the instruments used in this study and the preliminary processing of the observational data.

\subsection{Atmospheric Imaging Assembly (AIA)}
\label{sec:aia_instr}
The AIA instrument aboard {\it SDO} is a four-telescope array. It provides full-disk images of the sun and the inner corona up to 0.5 $R_\odot$ above the solar limb in seven EUV and three UV wavelength bands with high spatial ($\sim$1.2\arcsec) and temporal (12 s or 24 s) resolutions. From 11:29 UT to 15:28~UT on 2014 July 7, it captured a series of solar jets (see the movie) originating from a new emerging active region (NOAA 12114) at the south-east solar limb. 
The emissions measured in different EUV/UV channels are dominated by various spectral lines formed at different temperatures \citep{Lemen12}, i.e., AIA 1600 \AA\ (\civ, T$\approx$0.1 MK), 
304 \AA\  (\heii, T$\approx$0.05 MK), 
171 \AA (\feix, T$\approx$0.6 MK), 
AIA~193~{\AA} (\fexii/\fexxiv, T$\approx$1.6 \& 20~MK), 
211~{\AA} (\fexiv, T$\approx$2.0~MK), 335 \AA\ (\fexvi, T$\approx$2.5 MK), 94 \AA\ (\fexviii, T$\approx$6.3 MK), 131 \AA\ (\feviii/\fexxi, T$\approx$0.4 \& 10 MK). Thus, the multi-wavelength observations made by AIA can be used to diagnose the thermal structure of the corona \citep{ODwyer10,Del11,Del13}.

The AIA data were first downloaded from the {\it SDO} database in a format of level 1.0 and then were processed to level 1.5 via the standard AIA software ``aia\_prep.pro" which has been included in the Solar SoftWare (SSW) libraries. The EUV/UV intensities were normalized by the exposure time.

\subsection{Interface Region Imaging Spectrograph (IRIS)}
\label{sec:iris_instr}

{\it IRIS} provides simultaneous spectra and slit-jaw images (SJIs) of the solar atmosphere over a field-of-view (FOV) of up to 175\arcsec~$\times$175\arcsec\ with a temporal resolution of high up to 2 s and a spatial resolution of 0.33\arcsec$-$0.4\arcsec. 
The SJIs are taken in four different wavelengths which
are sensitive to  plasma emissions of different temperatures\citep{Dep14}, i.e.,  2830 {\AA} (\mgii wing, T $\approx$ 0.006 MK), 2796 {\AA} (\mgii k, T $\approx$ 0.01 MK), 1330 {\AA} (\cii, T $\approx$ 0.025 MK), and 1400 {\AA} (\siiv, T $\approx$ 0.08 MK). 
The spectra are obtained in far-UV (FUV: 1331.56$-$1358.40 \AA\ and 1390.00$-$1406.79 \AA) and near-UV (NUV: 2782.56$-$2833.89 \AA) passbands, containing some bright spectral lines (such as \mgii h 2803 \AA , \mgii k 2796 \AA , \cii 1334/1335 \AA, \siiv 1394/1403 \AA), with spectral resolutions of 26 m\AA\ in FUV and 53 m\AA\ in NUV.

On July 7, 2014,  {\it IRIS} pointed towards the eruption region (centered at $-$988\arcsec, $-$187\arcsec) with a maximum FOV of around 166\arcsec~$\times$~175\arcsec\ for SJIs.
Meanwhile, the spectra were obtained in a ``sit-and-stare'' mode, in which the {\it IRIS}  slit stays at a fixed location. The spectral dispersions are $\sim$51.92~m{\AA}/pixel in FUV wavebands and $\sim$50.92~m{\AA}/pixel in NUV wavebands.
We obtained {\it IRIS} data in a format of level 2 from the {\it IRIS} database, which are well prepared for easy use.

\subsection{Extreme-Ultraviolet Imager (EUVI)}

The EUVI instruments are a pair of identical telescopes onboard the twin {\it STEREO} spacecraft (one is behind and the other is ahead of the earth). They record images of the sun and the lower corona from two vantage points in four EUV channels (171 \AA, 195 \AA, 284 \AA, 304 \AA) whose emissions are dominated by four spectral lines (\feix, \fexii, \fexv and \heii) that span a temperature range of 0.05--20 MK. The EUVI's detectors have a FOV out to 1.7 $R_\odot$ from the sun center with a spatial resolution of about 3.2\arcsec.  
They have two observation modes: the synoptic mode and the campaign mode. Generally, They are operated in the synoptic mode, but for some special campaigns, it will be changed to the campaign mode which has fast cadences (two weeks in May 2007). 
During the event under study, only the EUVI onboard the {\it STEREO}-Behind was working with a cadence of 5 minutes in 195 \AA\ and of 120 minutes in the other wavelengths.
Although EUVI has very few observations, it did capture the largest jet, as shown in Fig.~\ref{snap}(c).
Meanwhile, the separation angle between {\it STEREO}-B and {\it SDO} is about 163$^{\circ}$, allowing us to make a 3D reconstruction of the jet. 

\subsection{Radio Spectrometers}

The related radio dynamic spectra were well recorded by  {\it WIND}/WAVES, {\it STEREO}-B/WAVES (SWAVES) and Nancay Decameter Array (NDA). {\it WIND}/WAVES and SWAVES are  
space-based spectrometers, which measure the solar radio emissions from the outer space in a low frequency range ({\it WIND}/WAVES: 0.02$-$13.825 MHz, SWAVES: 0.02$-$16.025 MHz) with a time cadence of about 1~minute. While NDA measures the radio emissions at earth in a higher frequency range of 10$-$80 MHz  with a higher temporal resolution of about 1~s. 

\section{Data analysis and results}

\subsection{Overview and kinematics of solar jets}

\begin{figure*}
\epsscale{.8}   \plotone{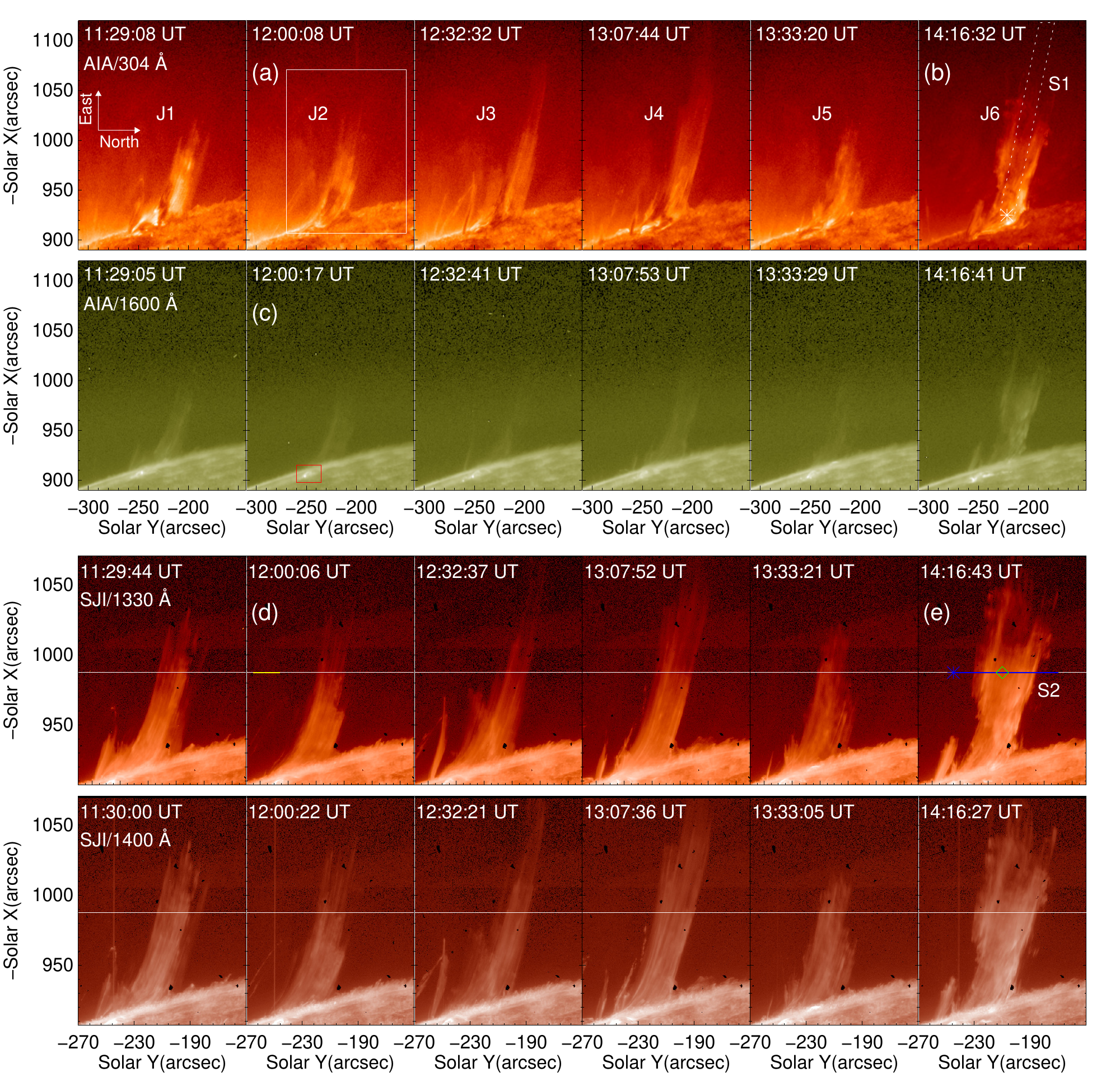} 
\caption{
The {\it SDO}/AIA 304 {\AA} (first row), 1600 {\AA} (second row), {\it IRIS}/SJI 1330 {\AA} (third row) and 1400 {\AA} (fourth row) images of six recurrent jets observed at 11:29 UT (J1), 12:00 UT (J2), 12:32 UT (J3), 13:07 UT (J4), 13:33 UT (J5) and 14:16 UT (J6) on July 7, 2014.
Images have been rotated so that the solar east is up.
SJI has a smaller field of view (FOV), indicated by the white box in panel (a). 
The white line on the SJI images represents the {\it IRIS} slit position.
The rectangle in panel (b) and the blue line in panel (e) represent two artificial slits (S1 and S2) where the time-distance plots were obtained. The ``$\ast$'' symbols indicate the start points of S1 and S2. Note that S2 is part of the {\it IRIS} slit. The green ``$\diamond$" represents the intersection point between the jet axis and the {\it IRIS} slit. The yellow line in panel (d) denotes a relatively quiet region where the rest wavelengths of {\it IRIS} spectra were determined.}
\label{image}
\end{figure*}

The movie (jet0707.mp4) generated from the {\it SDO}/AIA images and {\it IRIS} SJIs show a series of solar jets erupted from the same location close to the east limb of the Sun between 11:29:36 UT and 15:28:10~UT on July 7, 2014. These jets propagated in almost the same direction towards north-east and exhibited clear rotating motions as they propagated outward. 
Fig. \ref{image} presents snapshots of six of the jets (labelled as J1, J2, J3, J4, J5, and J6) in AIA 304 \AA~(first row),  AIA 1600 \AA~(second row), SJI 1330 \AA~(third row) and SJI 1400 \AA~(fourth row).
Observations show that these jets have a peculiar shape, which might be interpreted as a twisted flux rope.
They can be best seen in AIA 304 {\AA} images, SJIs 1330 {\AA} and 1400 {\AA}, and also in AIA 1600 {\AA} images and each of them exhibits a brightening point at their footpoints.
The first five jets (J1--J5) were small and had a lifetime less than 30 minutes. The last jet (J6) was much larger and lasted for about one hour.
Note that the FOV of SJIs (indicated by the white box in Fig.~\ref{image}(a)) is smaller than that of AIA.

\begin{figure*}
\epsscale{1.0} \plotone{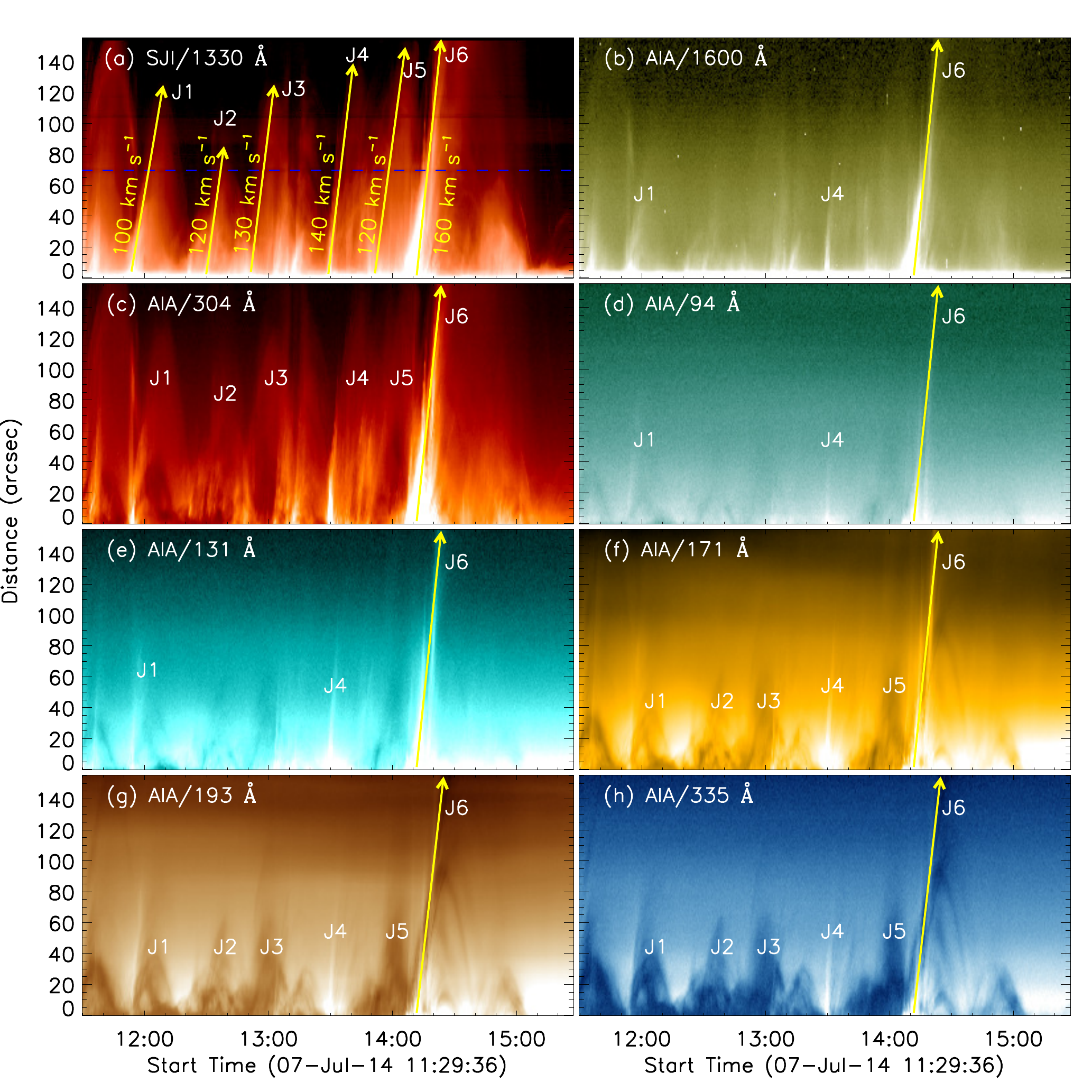} 
\caption{
Time-distance plots for S1, obtained using images taken in SJI 1440 {\AA} (a), AIA 1600 {\AA} (b), AIA 304 {\AA} (c), AIA 94 {\AA} (d), AIA 131 {\AA} (e), AIA 171 {\AA} (f), AIA 193 {\AA} (g) and AIA 335 {\AA} (h).
The yellow arrows illustrate the temporal evolutions of the six selected jets, and the dashed blue line in panel (a) represents the {\it IRIS} slit position. }
\label{slit1}
\end{figure*}

The white rectangle overplotted in Fig.~\ref{image} (b) represents an artificial slit (S1) with a certain width. The ``$\ast$'' symbol marks its starting point. 
Along S1, we perform a study of the jet kinematics by applying the time-distance (TD) technique on images obtained at various UV/EUV passbands.
Fig.~\ref{slit1} shows the TD plots in SJI 1330 {\AA}, AIA 1600 {\AA}, AIA 304 {\AA}, AIA 94 {\AA}, AIA 131 {\AA}, AIA 171 {\AA}, AIA 193 {\AA} and AIA 335 {\AA}, where the horizontal axis represents time and the vertical axis represents the propagation distance of the jets. The TD plots in SJI 1400 {\AA} and AIA 211 {\AA} passbands are shown in Fig.~\ref{spec1} (c) and Fig.~\ref{spec1} (d). 
As we can see, the jets can be clearly seen in low-temperature channels such as SJI 1330 {\AA}, SJI 1400 {\AA} and AIA 304 {\AA},  while in high-temperature channels such as AIA~94~{\AA}, 131~{\AA}, 171~{\AA}, 193~{\AA}, 211~{\AA} and 335~{\AA}, most of the jets appear as dark structures, suggesting that they are cool-material dominant (the temperature was supposed to be less than 1 MK considering the response temperatures of these wavelengths, see Sec. \ref{sec:aia_instr} and \ref{sec:iris_instr} for details).
In case of J1, J4 and J6, they also contain a thread-like structure which are bright in almost all UV and EUV emission lines, indicative of multi-temperature structures.
The multi-temperature material precedes the cool jet material, which might be heated by the superthermal particles accelerated during the magnetic reconnection process. The existence of superthermal particles can be verified by the observed type III radio bursts, as shown in Fig.~\ref{flux_typeIII}.

The yellow arrows in Fig.~\ref{slit1} (a) show the motions of the six jets. Note that there are also some jets which are not considered here due to their weak signal or incomplete observations of the instrument.
Using the SJI 1330 {\AA} TD plots, we calculated the average apparent velocities of the jets, which show that J6 has a larger velocity ($\sim$160 km s$^{-1}$) than the other jets (i.e., 100 km s$^{-1}$  for J1, 120 km s$^{-1}$ for J2, 130 km s$^{-1}$ for J3, 140 km s$^{-1}$ for J4 and 120 km s$^{-1}$ for J5) and propagated to a much higher height.
For each jet, we also found some ejected materials falling back to the solar surface with various speeds, especially for J6, whose falling speed ranges from $\sim$40~km~s$^{-1}$ to $\sim$170~km~s$^{-1}$, as shown in Fig.~\ref{spec1} (c).

\begin{figure*}
\epsscale{1.0} \plotone{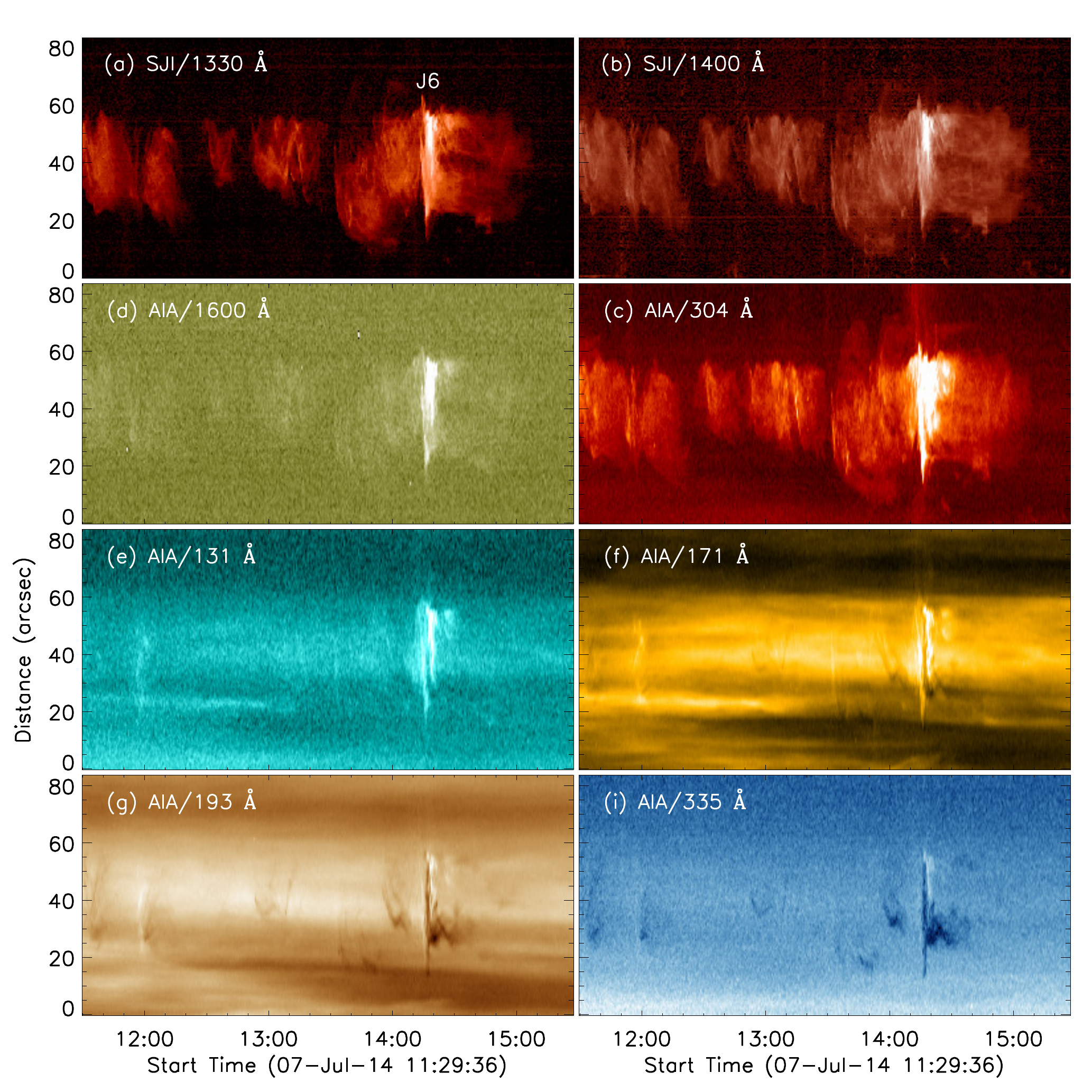} 
\caption{Time-distance plots along S2, obtained from images taken in SJI 1330 {\AA} (a), 1400 {\AA} (b), AIA 304 {\AA} (c), AIA 1600 {\AA} (d), AIA 131 {\AA} (e), AIA 171 {\AA} (f), AIA 193 {\AA} (g) and AIA 335 {\AA} (i).}
\label{slit2}
\end{figure*}

The overplotted white lines on the SJI images in Fig.~\ref{image} show the position of the {\it IRIS} slit, which was almost perpendicular to the jet axis.
Fig.~\ref{slit2} shows the TD plots in part of the {\it IRIS}  slit (S2, blue line in Fig.~\ref{image}(e)) in multiple {\it IRIS}/SJI and {\it SDO}/AIA passbands,  from which the width of the six jets can be estimated, i.e., 23 Mm for J1, 15 Mm for J2, 19 Mm for J3, 40 Mm for J4, J5 and J6.

\subsection{Spectroscopic observations of the jets}
\label{sec:spectral_study}

 As mentioned above, during the recurrent activity of the jets, the {\it IRIS} slit was operated in a ``sit-and-stare'' mode and was just located above the active region, oriented almost perpendicular to the jet axis, as shown by the white lines on the SJIs in Fig.~\ref{image}. 
This is a very fortunate case, which enables us to perform a spectral study of these transient and short-duration jet events.

\begin{figure*}
\epsscale{1.0} \plotone{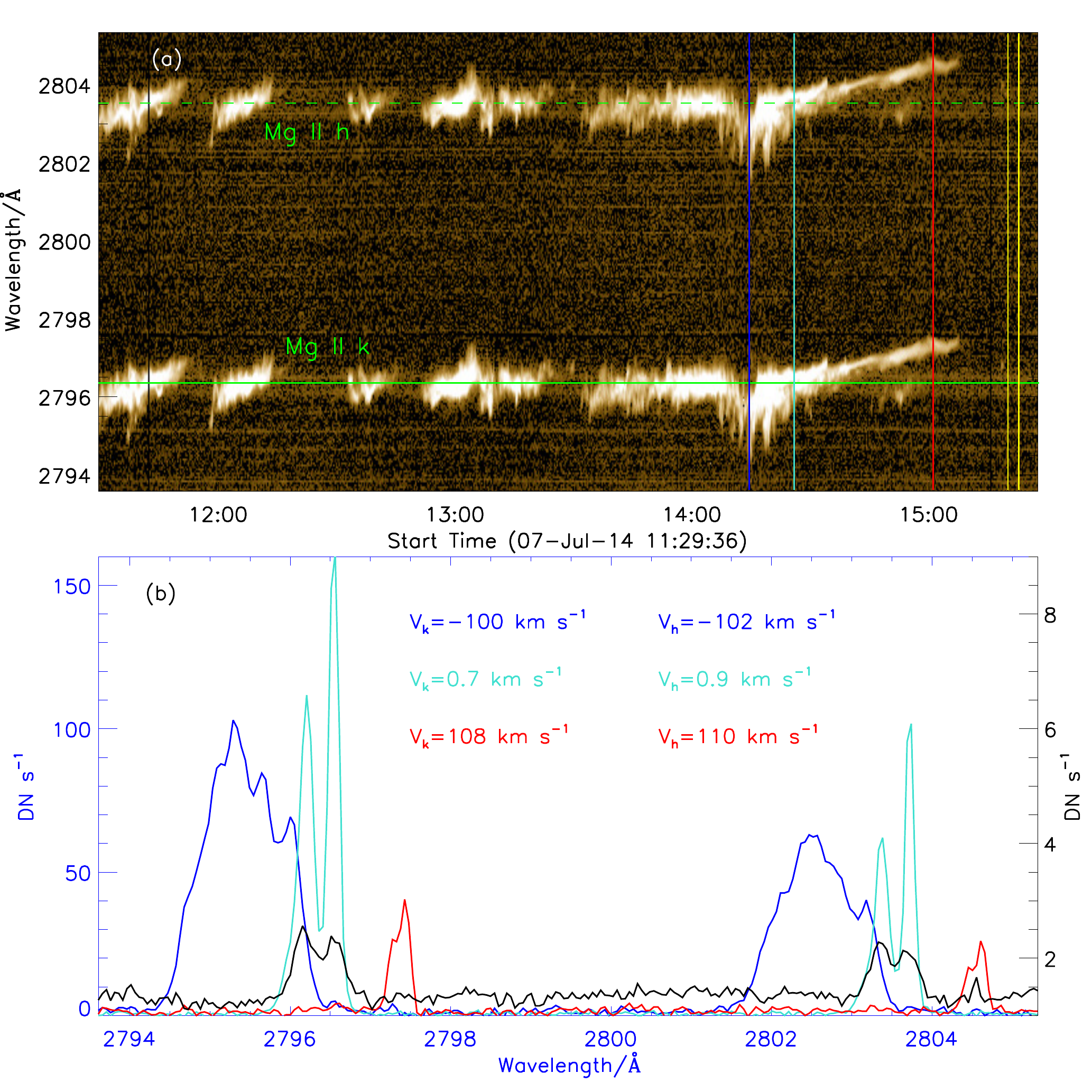} 
\caption{Panel (a): Time-wavelength plot of the resonance \mgii lines at the intersection point (indicated by the green ``$\diamond$" in Fig.~\ref{image} (e)) between the jet axis and the {\it IRIS} slit.
The two horizontal green lines outline the rest wavelengths of \mgii~k 2796~{\AA} and \mgii~h 2803~{\AA}. Panel (b): Line profiles of \mgii h \& k at three times indicated by the blue, cyan and red vertical lines in panel (a). The black line represents the quiet spectrum calculated during the time interval of 15:20:02~UT$-$15:22:45~UT indicated by the two  yellow vertical lines in panel (a).} 
\label{spec2}
\end{figure*}

Fig.~\ref{spec2}~(a) presents the temporal evolution of the line spectrum at the intersection point (indicated by the green $\diamond$ in Fig.~\ref{image} (e)) between the jet axis and the {\it IRIS} slit.
The vertical axis represents the wavelength values. 
As can be  seen, {\it IRIS} observed the strong emissions of the resonance \mgii lines (\mgii~h $\&$~k) during the jet eruptions. 
Note that the vertical black line at about 11:50~UT is caused by a data gap. The horizontal green lines indicate the rest wavelengths of \mgii~h (2803.53~{\AA}) $\&$~k (2796.35~{\AA}), which are obtained by averaging the spectra from a quiet region (indicated by the yellow bar in Fig.~\ref{image}(d)) during a quiet time interval from 15:20:02~UT to 15:22:45~UT (denoted by the two vertical yellow lines in Fig.~\ref{spec2} (a)).

To display the spectral information of the largest jet `J6' in more details, we plotted the line profiles (as shown in Fig.~\ref{spec2} (b)) at three different times (indicated by the vertical colorful lines in Fig.~\ref{spec2} (a)).
The black curve illustrates the quiet spectrum of the resonance \mgii lines described above. The blue, cyan and red curves show the line profiles observed at the start time (14:14:26~UT), the intermediate time (14:25:49~UT) and the end time (15:01:03~UT) of J6, respectively.

As can be seen,  both \mgii~h~$\&$~k lines deviate from a Gaussian shape, which exhibit a central reversal in the line core. This means that the frequently used Gaussian-fitting method may not be suitable here \citep{Leenaarts13a,Leenaarts13b,Cheng15}. 
In this work, we apply the moment method to estimate their line intensity and line center. In combination with the rest wavelengths as described above, the Doppler velocities can be calculated \citep[see][]{Li17}. 
Our calculations show that the largest jet J6 exhibits strong blue-shifted Doppler velocity at the beginning (about $-$100~km~s$^{-1}$ in \mgii~k and $-$102~km~s$^{-1}$ in \mgii~h at 14:14:26~UT). Then the Doppler velocity gradually decreases to zero and even changes to red-shifted values (about 108~km~s$^{-1}$ in \mgii~k and 110~km~s$^{-1}$ in \mgii~h at 15:01:03~UT). 

\begin{figure*}
\epsscale{1.0} \plotone{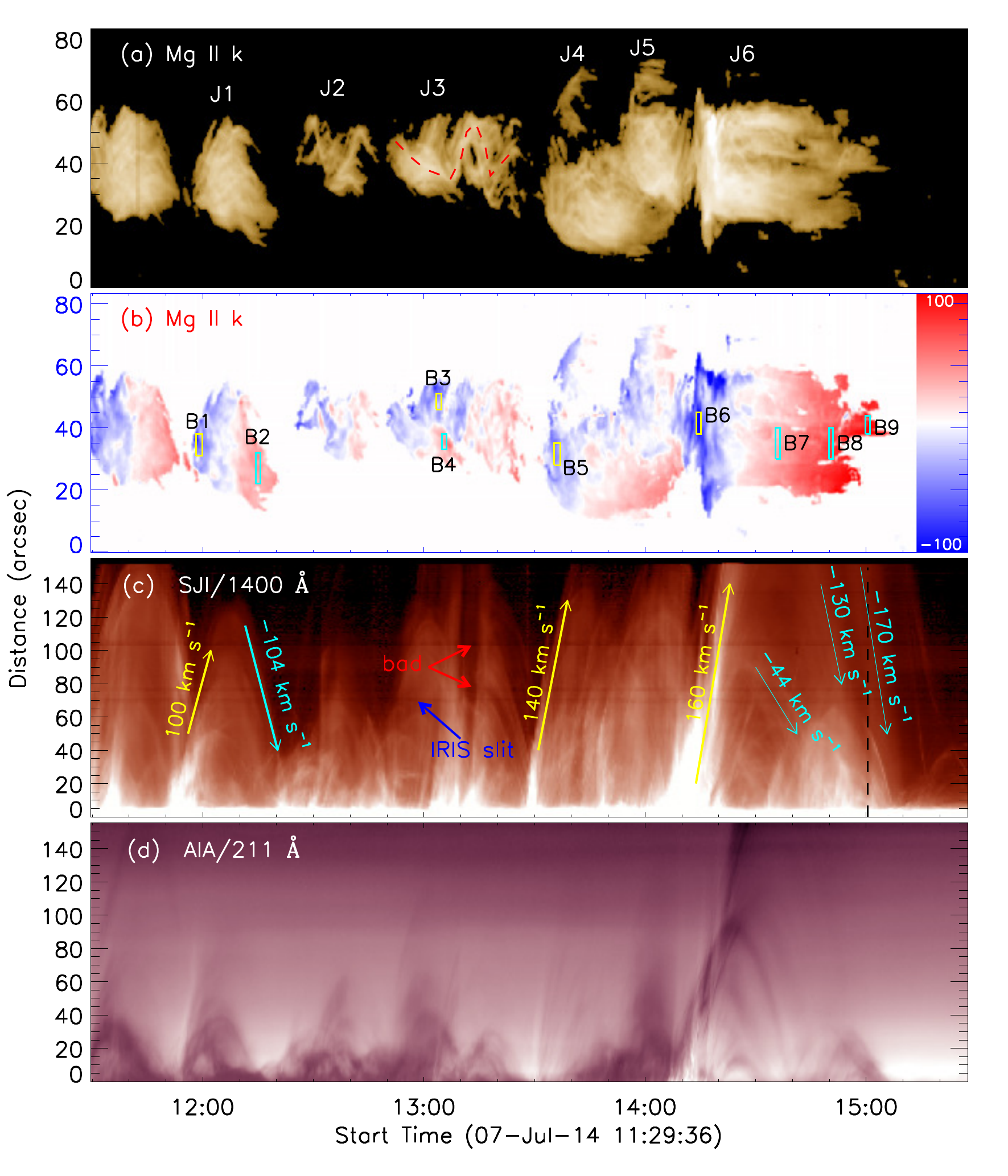} 
\caption{
Panel (a): temporal evolution of \mgii~k intensity along S2. The red curve indicates the rotating motions of J3.
Panel (b): temporal evolution of \mgii~k Doppler shift calculated by the moment method. 
The nine boxes represent nine subregions, as denoted by ``B1"--``B9". 
Panel (c) and panel (d) show similar time-distance plots as in Fig.~\ref{slit1}, but in SJI 1400 {\AA} (c) and AIA 211 {\AA} (d). The yellow and cyan arrows indicate upward and downward motions of the jet materials, respectively. The black horizontal lines in panel (c) are caused by some bad pixels at the {\it IRIS} slit.} 
\label{spec1}
\end{figure*}

We note that the \mgii~k 2796~{\AA}  and \mgii~h 2803~{\AA} lines exhibit similar bahavious and both of them show a higher signal-to-noise ratio than the other lines (not shown here). To avoid repetition, we hereafter use the \mgii~k 2796~{\AA} line to make the following Doppler velocity calculations.
Fig.~\ref{spec1} (a) displays the temporal evolution of the \mgii~k line intensity along the {\it IRIS} slit, which displays a similar evolution as those measured in SJIs 1330 {\AA}, SJIs 1400 {\AA} and AIA 304 {\AA} images.
All the six selected jets were observed in the \mgii~k line and in case of J2 and J3, we found some sine-like tracks as indicated by the red dashed lines.
Considering the {\it IRIS} slit was oriented nearly perpendicular to the jet axis (Fig.~\ref{image}), these sine-like tracks might represent the rotating motions of the jets around their axes \citep{Liu14}.
The rotation periods are estimated to be around 20 minutes, which is very close to that reported by \cite{Liu14}.
Assuming that the jets rotate in a cylindrical trajectory, their rotating radius was estimated to be $\sim$25\arcsec\ ($\sim$1.8$\times$10$^4$ km). Based on this, the rotational speeds of the jets can be estimated to be $\sim$34 km s$^{-1}$.

Fig.~\ref{spec1} (b) presents the corresponding Doppler velocity evolution map of the \mgii~k line. It shows both blue and red Doppler shifts. 
On a close comparison of the Doppler shifts with the POS velocities measured from the TD plots in SJI 1400 {\AA} (c) and  AIA 211 {\AA} (d), we found that the blue shifts are associated with the radial outflows of the jet materials while the red shifts are related to their downward fallings. 
Considering that the line shifts are caused by the line of sight (LOS) motions of the jet material, an inclination angle ($\theta$) that the jets deviate from the plane of sky (POS) is expected.
Meanwhile, the jets J2$-$J5 also exhibit some oppositely shifted components on their two sides, indicative of rotating motions along their axes, as can be seen from Fig.~\ref{spec1} (a) and the movie jet0707.mp4.

To make a more detailed analysis of the Doppler velocities, we selected nine subregions (denoted by B1$-$B9)  marked by the yellow (blue-shifted regions) and cyan (red-shifted regions) boxes in Fig.~\ref{spec1} (b).
The Doppler velocities within these nine subregions were given in the second row of Table \ref{tab1}.
Note that B3 and B4 represent two subregions on two sides of J3, where oppositely Doppler shifts were observed.
For the other subregions (B1$-$B2 and B5$-$B9), we distinguish their corresponding POS motions from the SJI 1400 {\AA} TD plot, shown by the yellow (upward motions) and cyan (downward motions) arrows in Fig.~\ref{spec1}(c), and their velocities were listed in the third row of Table \ref{tab1}. 
Making use of the Doppler velocities and the POS velocities, the three-dimentional (3D) velocities and the inclination angles ($\theta$) of the jets can be roughly estimated, as shown in the fourth and fifth rows of Table \ref{tab1}, respectively.

\begin{table*}[]
\centering
\caption{Velocities of recurrent jets for the nine subregions (B1-B9) from the Doppler velocity map.}
\label{tab1}
\begin{tabular}{@{}c|c|ccccccccc@{}}

\toprule
\multicolumn{2}{l}{} & B1 & B2 & B3 & B4 & B5 & B6 & B7 & B8 & B9 \\ \midrule
\multirow{4}{*}{} & Doppler\tablenotemark{1} & -55$\pm$16 & 36$\pm$5 & -32$\pm$9 & 31$\pm$7  & -38$\pm$7 & -92$\pm$17 & 28$\pm$5 & 69$\pm$6 & 88$\pm$15 \\
       {Velocity} & POS\tablenotemark{2} & 100  & -104 & - & - & 140 & 160 & -44 & -130 & -170 \\
{(km s$^{-1}$)} & 3D\tablenotemark{3} &115$\pm$8  & -110$\pm$2 & - & - & 145$\pm$2 & 185$\pm$8 & -52$\pm$3 & -147$\pm$3 & -192$\pm$7 \\
   & Angle ($\theta$)\tablenotemark{4} & 28$\pm$7 & 19$\pm$3 & - & - & 15$\pm$3 & 30$\pm$5 & 32$\pm$5 & 28$\pm$2 & 27$\pm$4 \\
\hline
\bottomrule  
\end{tabular}
\tablenotetext{1}{The Doppler velocities calculated using the {\it IRIS} \mgii~k line observations.}
\tablenotetext{2}{The plane-of-sky velocities obtained from the SJI 1400 {\AA} images.}
\tablenotetext{3}{The three-dimension (3D) velocities calculated by combining the Doppler and  plane-of-sky velocities.}
\tablenotetext{4}{The inclination angles that the jets deviate from the plane-of-sky of the observer.}
\end{table*}

\subsection{3D reconstruction of the largest jet (J6)}

\begin{figure*}
\epsscale{1.0} \plotone{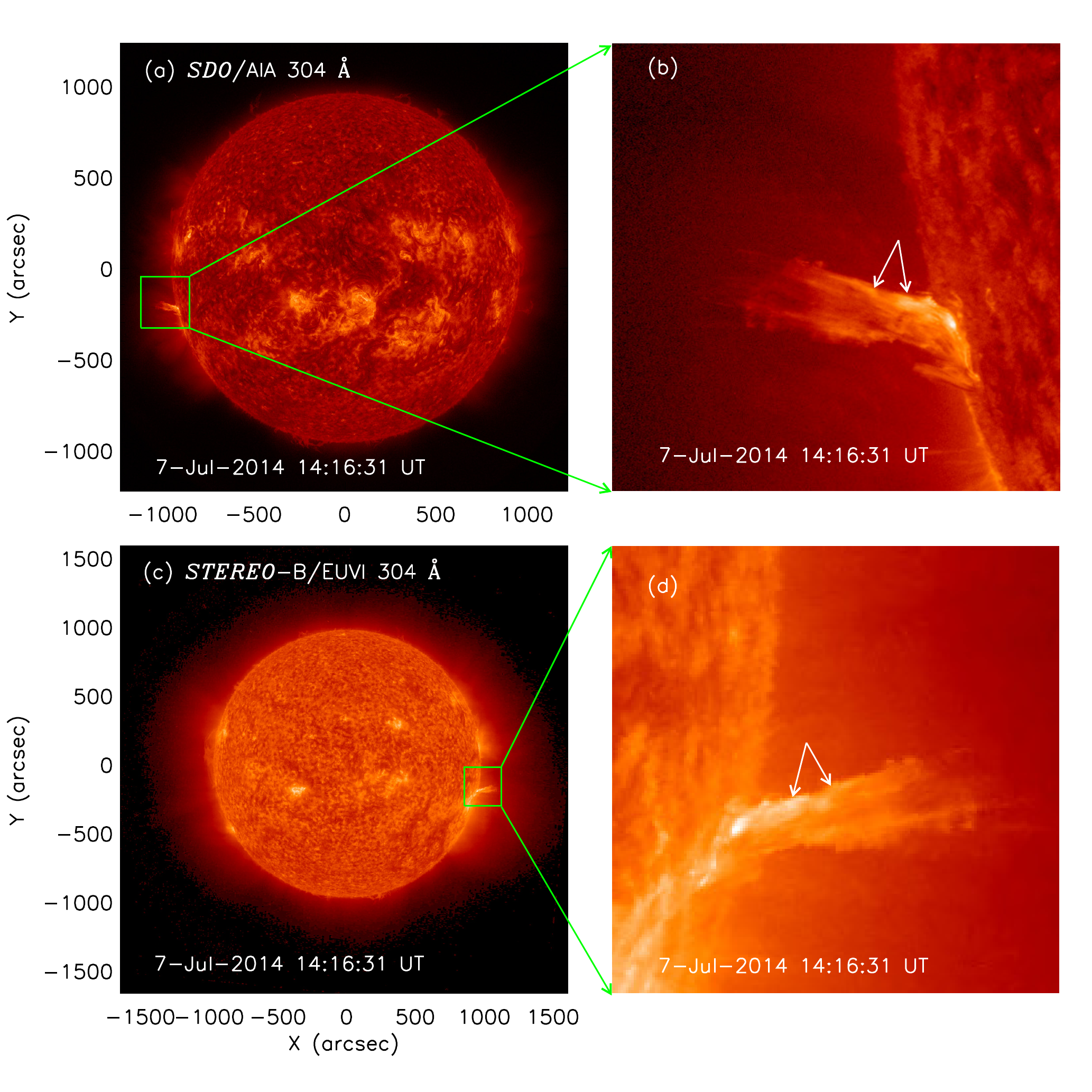} \caption{Joint observations of the solar jet at about 14:16~UT on 2014 July 7 in 304~{\AA} by {\it SDO}/AIA (a) and {\it STEREO}-B/EUVI (c) from two perspectives. Panels (b) and (d) show the corresponding zoom-in views (indicated by green boxes in the left-hand column) of the jet. The white arrows refer to two blob-like structures.} \label{snap}
\end{figure*}

The jets were also captured by the {\it STEREO}-B/EUVI from another perspective, but due to the low cadence of EUVI data, only one snapshot in 304 {\AA} was taken during the rising of the largest jet (J6) at 14:16 UT.
Fig.~\ref{snap} presents images of J6 obtained by {\it SDO}/AIA (a) and {\it STEREO}-B/EUVI (c). 
The separation angle between these two telescopes is about 163$^\circ$.
Panels (b) and (d) show the zoomed-in views, from which some bright and compact features (EUV blobs, indicated by white arrows, see \citealt{Zhang14b}) at the edge of the jet were identified. 
The joint observations from these two perspectives enable us to perform a 3D reconstruction of the jet, which can give information about the inclination angle that the jet makes with the POS.

The 3D reconstruction was performed using the the Interactive Data Language (IDL) procedure
 ``scc\_measure.pro" provided in the SSW package.
This routine contains an application widget which enables us to select a feature point of interest with the cursor in the image taken from one perspective ({\it SDO}/AIA image in our case). 
Then a line that represents the LOS from the AIA perspective is displayed in the image taken from another perspective ({\it STEREO}-B/EUVI image in our case).
Along this line, we then select the point with the same feature.
By this way, the 3D coordinates of the selected feature are determined as the Stonyhurst heliographic longitude and latitude as well as the radial distance in units of solar radii. 
Repeating the same operation, the 3D coordinates for all the feature points of the jet can be determined.

\begin{figure*}
\epsscale{1.0} \plotone{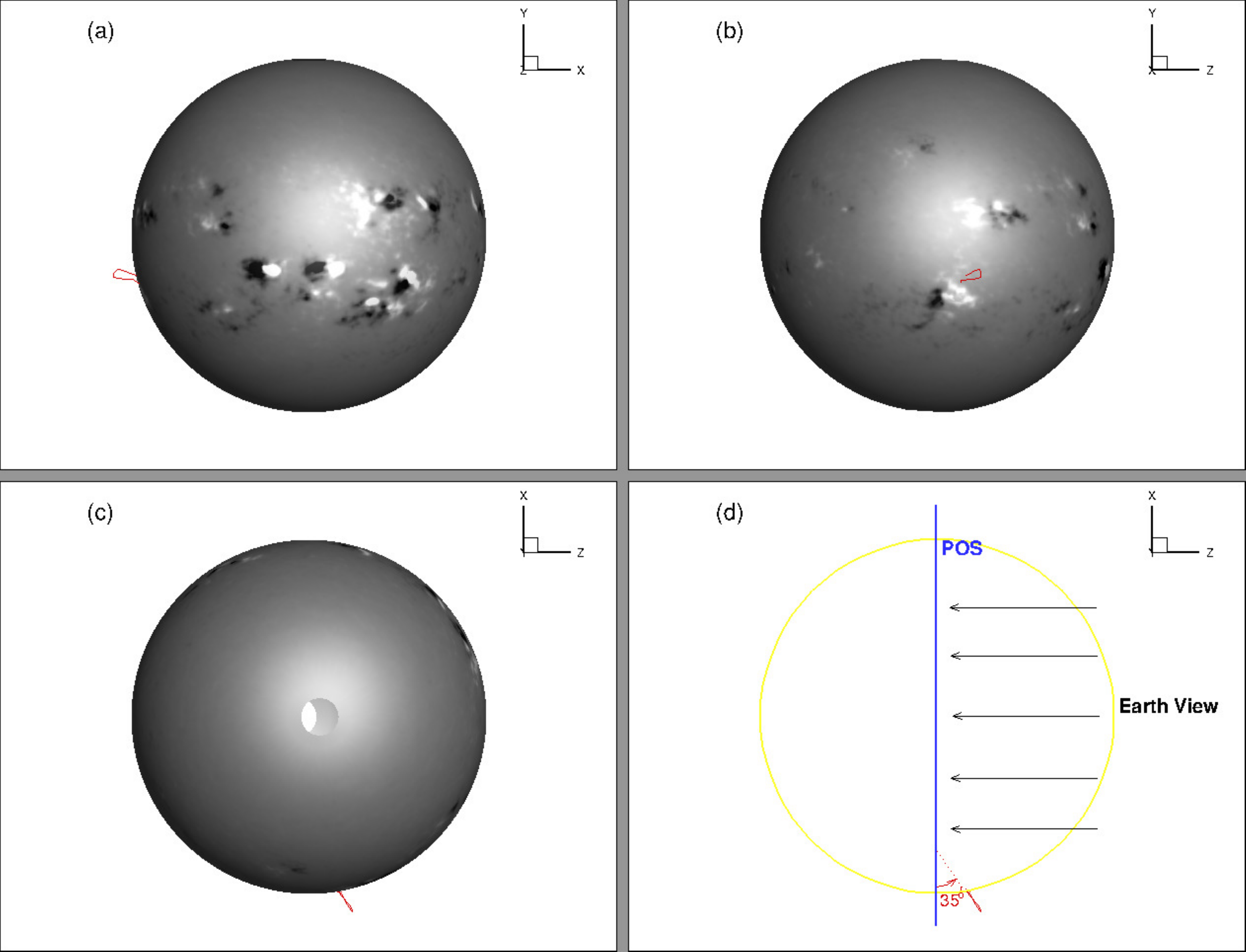} 
\caption{
3D reconstruction of the largest jet (red curve) as viewed along different orientations in the heliocentric-cartesian coordinate system. Panels (a) and (c) show two lateral views while panel (b) shows a top view. 
The grey sphere represents the GONG synoptic magnetogram of the Carrington Rotation (CR) 2152, where the white and black colors show the positive and negative magnetic field, respectively. 
The empty spaces in the synoptic magnetogram refer to the high-latitude regions where the measurements of the magnetic field are missing. Panel (d) is a cartoon showing the incination angle of the jet from the plane-of-sky (POS) of Earth.
Earth is in the positive Z direction. The plane containing the X and Y axes defines the POS.} \label{jet_3D}
\end{figure*}

The Stonyhurst coordinates can be easily converted into the heliocentric-cartesian coordinates in which the Z-axis is defined to be parallel to the observer-Sun line, pointing toward the observer. 
The X- and Y-axes are perpendicular to each other and defines the POS of the observer. Positive Y-axis is north and positive X-axis is west. Their rotations follow the right-hand rule (for a complete review of the coordinate systems, please refer to \citealt{Thompson2006}).
In Fig.~\ref{jet_3D}, we display the 3D reconstruction result of the largest jet J6 from different viewing angles in heliocentric-cartesian coordinate system. The gray sphere represents the  GONG synoptic magnetogram for Carrington Rotation (CR) 2152. Panel (a) shows the geometric configuration of the jet seen from Earth (along Z-axis). Panel (b) shows the top view of the active region (AR 12114) where the jets erupted (along X-axis), from which we can see that the J6 locates at the north-west edge of the active region and is close to the positive magnetic polarity region. Panel (c) shows a side view of the jet (along Y-axis), from which the inclination angle of the jet deviating from the POS can be estimated. Panel (d) is a cartoon showing the same viewing angle as Panel (c) but without the magnetogram overplotted.
The inclination angle of J6 we calculated here ($\sim$35$^\circ$) is very close to the value that was calculated through the imaging and spectral analysis method (B6-B9 in Table.~\ref{tab1}), further verifying the reliability of the two methods.

\subsection{Relationship to other solar eruptions}

\begin{figure*}
\epsscale{1.0} \plotone{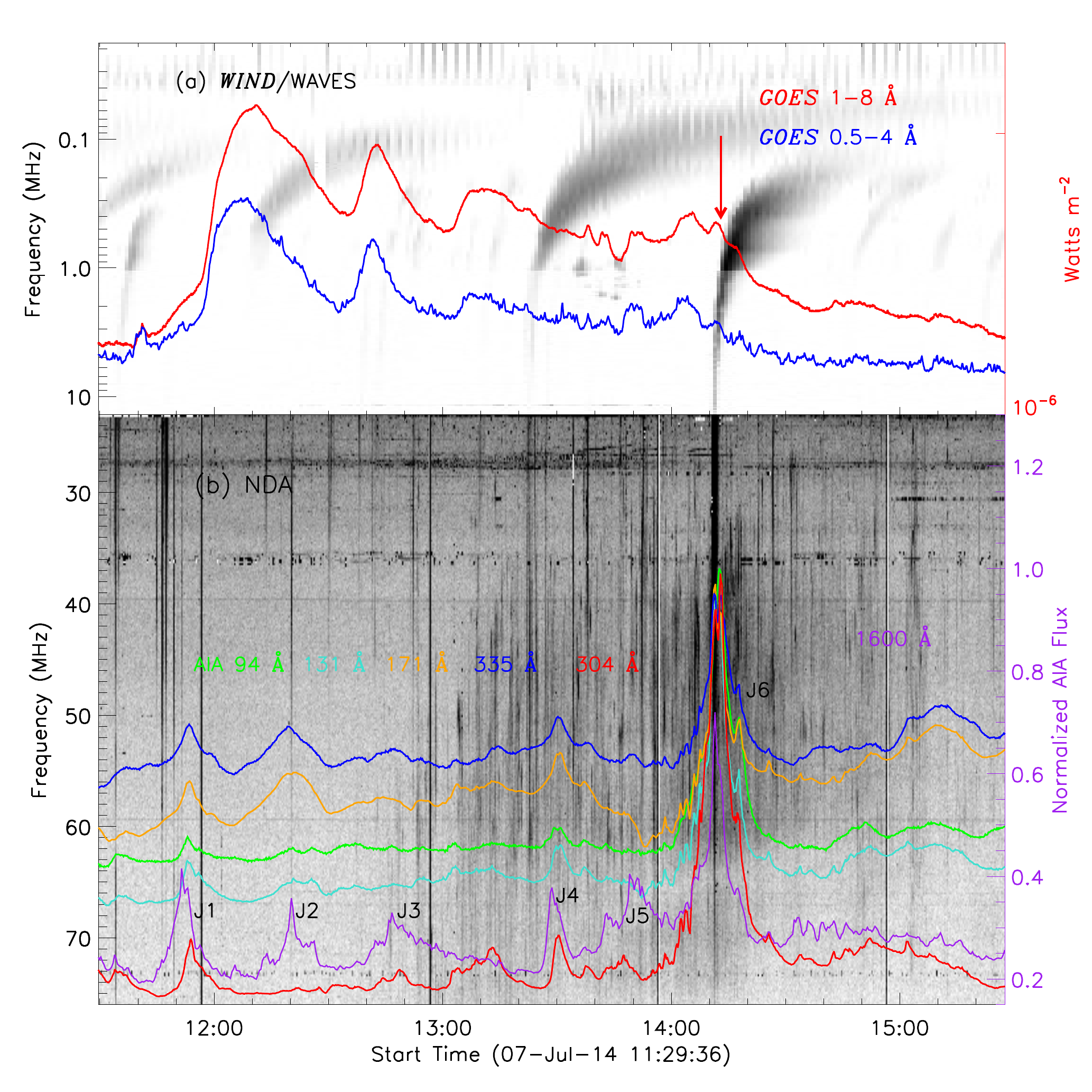} \caption{Panel~(a): SXR fluxes
measured by {\it GOES} in 1$-$8~{\AA} (red) and 0.5$-$4~{\AA} (blue), with the dynamic spectra observed by {\it WIND}/WAVES as its context.
Panel~(b): the local UV/EUV light curves integrated over the active region shown by the red box in Fig.~\ref{image} (c). The context is the dynamic spectra detected by NDA. 
J1--J6 indicate the six jets as shown in Fig.~\ref{image}.} \label{flux_typeIII}
\end{figure*}

Fig.~\ref{flux_typeIII} (a) shows the full solar disk SXR fluxes measured by {\it GOES} in 1$-$8~{\AA} (red) and 0.5$-$4~{\AA} (blue), with the dynamic spectra observed by {\it WIND}/WAVES as its context. The {\it GOES} SXR light curves exhibit a few peaks, which could be considered as a few C-class flares according to their peak intensities in 1$-$8~{\AA}. To locate these flares, space-resolved observations such as the full-disk images from {\it SDO}/AIA are used. 

Fig.~\ref{flux_typeIII} (b) shows the normalized light curves integrated over the jet-related active region (outlined by the red box in Fig.~\ref{image} (c)) in AIA's six passbands. The context is the radio dynamic spectra detected by NDA, from which we see a number of type III radio bursts.
Note that the light curves have been shifted for a better display. 
There are notable differences between these light curves observed at different passbands and only the light curve at 1600 \AA\ shows clear responses to each of the jets under study (labelled by J1$-$J6).
Comparing the local-region light curves with the full-disk SXR flux, we find that only the last {\it GOES} flare (C1.6, indicated by the red arrow) comes from the active region and the other flares that occurred earlier are considered to come from some other active regions after a careful examination (see an accompany paper by Li at al, 2019).

The C1.6 flare and the largest jet (J6) occurred almost simultaneously, also accompanied by a strong type III radio burst as recorded by both {\it WIND}/WAVES and NDA. This implies that the ``J6" jet was probably triggered by the C1.6 flare, during which a large number of nonthermal electrons are accelerated and ejected.
Moreover, in a larger FOV, a weak CME event was observed by the white-light coronagraph (C2) onboard {\it SOHO}/LASCO. 
The CME erupted just after the J6 jet (at about 14:36 UT), originated from the same source region, and propagated along almost the same direction (position angle: 103$^\circ$), which lead us to a possibility that the CME was triggered by the J6 jet.
Observations of the C1.6 flare, type III radio burst and the CME provide evidence of magnetic reconnection, which results in strong emission enhancements in all the AIA passbands as shown in Fig.~\ref{flux_typeIII} (b). 
The first five AIA emission enhancements are relatively weak and did not cause clear {\it GOES} flares, which might be considered as EUV bright points (EBPs), as shown in panel~(b). 
As can be seen, some EBPs (such as J1 and J4) are significantly enhanced in all of the AIA channels, while some others ( such as J2, J3 and J5) only show visible emission enhancements in some of the AIA channels.  
All the EBPs are found to be well correlated with the above-discussed five jets (J1$-$J5) but without CMEs or radio bursts accompanied, which may indicate that a small-scale magnetic reconnection is at work.  

\section{Summary and discussion}

In this work, we investigated multi-spacecraft observations of a series of recurrent jets during the time interval of 11:29:36--15:28:10~UT on 2014 July 7. 
The jets were found to be homologous and originated from the northwest periphery of a new emerging active region NOAA AR 12114. 
The multi-wavelength observations show that some of the jets (J2, J3 and J5) are bright in low-temperature channels (e.g., SJI~1330~{\AA} \&\ 1400~{\AA} and AIA~304~{\AA} \&\ 1600~{\AA}), but dark in high-temperature channels 
(e.g., AIA~131~{\AA}, 171~{\AA}, 193~{\AA}, 211~{\AA}, 335~{\AA}, 94 {\AA}), indicative of their low-temperature characteristic (less than 1~MK).
For the other jets (J1, J4 and J6), they contain multi-temperature plasma that was bright in almost all AIA and SJI channels.
The multi-temperature plasma appears as a thread-like structure and precedes the associated cool jets in the TD plots.

\cite{Yokoyama95}  made a two-dimensional numerical modeling of coronal jets and proposed that both cool and hot plasmas are expected to be accelerated during the magnetic reconnection between the emerging flux and open magnetic fields.  
\cite{Nishizuka08} extended the model of \cite{Yokoyama95} by considering more realistic initial conditions for the corona and reported a temperature of $\sim$10$^5$ K for the cool jets and a temperature of more than $5 \times 10^6-10^7$ K for the hot jets, which are also consistent with the temperatures we found here.
Additionally, \cite{Shen12,Shen17} found that the cool components of the jet are actually some minifilaments which erupted from  the jet base, while \cite{Jiang07} proposed that the cool plasma flows are possibly formed during the cooling process of the hot ones.
Therefore,  the generation of the cool and hot plasma flows in solar jets is still an open question.

The TD analysis in SJI 1330 {\AA} shows that the average POS velocity of the jets ranges from 100 km s$^{-1}$ to 160 km s$^{-1}$, which are comparable to those reported before \citep{Chandra2015,Mulay17}. 
After reaching the maximum height, some of the jet materials fell back to the solar surface with velocities varying from 44 km s$^{-1}$ to 170 km s$^{-1}$.
Moreover, the movie generated from SJIs and AIA images shows clear rotating motions of the jets as they propagated outward.

\cite{Tian08} found cool and hot components in a coronal bright point showing totally different orientations and interpreted them as a twist of the magnetic loop systems. 
If the twisted loop systems are commonly associated with bright points or jets, they may provide the initial magnetic field topology that leads to the rotating motions in jets.
\cite{Canfield96} observed rotating motions of H$\alpha$ surges and interpreted them as a consequence of the relaxation of stored magnetic twist. The same mechanism may also work in the jets discussed here. 
However, the origin of the twist is still not fully understood and there are still some issues to be addressed in the modeling of jets.

These jets were fortunately captured by the {\it IRIS} spectroscopic observations in multiple spectral lines  such as \mgii~h~$\&$~k, \cii~1335 \AA\ and 1336~{\AA}, and \siiv~1403~{\AA}, et al. 
All these spectral lines show that the Doppler velocities of the jets evolve continuously from blue shifts to red shifts with the radial motions changing from upward to downward. 
The Doppler velocity represents the LOS component of the jet motions. An examination of nine subregions (B1$-$B9) on the Doppler velocity evolution map reveals that the Doppler velocity of the jets ranges from 20 to 110 km s$^{-1}$. In combination with the POS velocities measured from TD plots, the 3D velocity and the inclination angle of the jets were estimated, i.e., $115 \pm 8$ km s$^{-1}$ ($28^\circ \pm 7^\circ$) for B1, $-110 \pm 2$ km s$^{-1}$ ($19^\circ \pm 3^\circ$) for B2, $145 \pm 2$ km s$^{-1}$ ($15^\circ \pm 3^\circ$) for B5, $185 \pm 8$ km s$^{-1}$ ($30^\circ \pm 5^\circ$) for B6, $-52 \pm 3$ km s$^{-1}$ ($32^\circ \pm 5^\circ$) for B7, $-147 \pm 3$ km s$^{-1}$ ($28^\circ \pm 2^\circ$) for B8, and $-192 \pm 7$ km s$^{-1}$ ($27^\circ \pm 4^\circ$) for B9.
B3 and B4 represent a pair of subregions where oppositely directed flows appeared (i.e., $-32 \pm 9$ km s$^{-1}$ for B3 and $31 \pm 7$ km s$^{-1}$ for B4), indicative of rotating motions along the jet axis.
For the subregions B2--B5, they involve both radial and rotating motions of the jets and probably due to the cancellation between these two kinds of motions, they exhibit relatively smaller Doppler shifts, and thus resulting in smaller inclination angles.  

The dual-perspective observations from {\it SDO}/AIA and {\it STEREO}-B/EUVI enable us to make a 3D reconstruction of the jet J6, which allows us to make a more accurate calculation of the speed and direction of the jet. It reveals that the jet propagated at a speed of $\sim$193 km s$^{-1}$ along a direction deviating from the POS of $\sim$35$^\circ$, which are well consistent with the results estimated from the imaging and spectral analysis.
For this work, we are  not able to make a detailed analysis on the 3D evolution of the jet structure due to the low cadence of {\it STEREO}-B data, although it captured the jets from another perspective.

Our observations show that the jet J6 was triggered by a C1.6 flare, while the others can only be associated with BPs.
It is well known that both the BPs and flares represent the  magnetic energy release through the magnetic reconnection process \citep[e.g.,][]{Masuda94,Aschwanden02,Zhang12,Ning14,Yan18}. 
This can be supported by type III radio bursts (signatures of non-thermal particles accelerated by magnetic reconnection) recorded  by the {\it WIND} and NDA instruments during the jet erutions.
Usually, the energy of BPs is released via the small-scale magnetic reconnection \citep{Solanki93},  which is too weak to cause the soft X-ray flares in the GOES flux curve. Alternatively, the absence of soft X-ray flares could also be due to the occulting of the solar disk since the jets lasted for several hours and the reconnection sites in the first five jets could locate behind the solar disk.
The driven energy of J6 is much stronger than its previous ones in terms of its maximum height, width, and emission intensity, which even leads to a weak CME as observed by LASCO/C2 in the outer corona. This agrees with previous findings that the jets can play an important role in the energy and mass input to the CMEs and solar winds \citep{Moore10,Raouafi16,Shen18c}. 
Our observational result also provides a good example supporting the study by \cite{Shen11b} in which the authors claimed that a CME can only be launched if the energy released in the low coronal exceeds some critical values, given the similar initial coronal conditions.

\acknowledgments  The authors thank the
referee for his/her valuable comments and suggestions. The data used in this paper are from {\it IRIS}, {\it SDO}, {\it STEREO}, {\it WIND}, {\it NDA}, and {\it GOES}. We appreciate for their open data use policy. This work is supported  by  the National Key R\&D Program of China (2018YFA0404200),  NSFC (grant Nos. 11522328, 11473070, 11427803, U1731241, 11603077, 11573072, 11873095 and 11973092),  the Youth Fund of Jiangsu (grant Nos. BK20161095, BK20171108), as well as CAS Strategic Pioneer Program on Space Science (grant Nos. XDA15010600, XDA15052200, XDA15320103, and XDA15320301). 
The Laboratory No. 2010DP173032.
L.F. also acknowledges the Youth Innovation Promotion Association for financial support. 
Y.L. is also supported by the CAS Pioneer Hundred Talents Program.

\end{document}